\documentclass [12pt]{article}
\usepackage{amsmath,amssymb,cite}

\usepackage{tabularx}
\usepackage[english]{babel}
\usepackage[dvips]{graphicx}
\usepackage{indentfirst}
\usepackage{epsfig}
\numberwithin{equation}{section}

\begin{document}

\title{
Topology Change From Quantum Instability of Gauge Theory on Fuzzy ${\rm CP}^2$}

\author{Djamel Dou$^{1,2}$\footnote{ddou@ictp.it}, Badis Ydri$^{3}$\footnote{ydri@physik.hu-berlin.de.}\\
$^{1}$ Dept of Physics and Astronomy, King
Saud University\\ P.O. Box 2455 Riyadh 11451, Saudi Arabia.\\
$^{2}$ Institute of Exact Science and Technology, University
Center of Eloued\\ Eloued, Algeria\\
$^{3}$Institut fur Physik, Humboldt-universitat zu Berlin\\
Newtonstr.15, D-12489 Berlin-Germany.}

\maketitle

\begin{abstract}

Many gauge theory models on fuzzy complex projective spaces  will
contain  a strong instability  in  the quantum field theory
leading to  topology change.
 This can be thought of  as due to the interaction between spacetime via its noncommutativity and the fields (matrices) and it is related to the
  perturbative UV-IR mixing. We work out in detail the example of fuzzy ${\bf CP}^2$ and discuss at the level of the phase diagram the quantum
  transitions between the $3$ spaces ( spacetimes) ${\bf CP}^2$, ${\bf S}^2$ and the $0-$dimensional space consisting of a single point $\{0\}$.

\end{abstract}


 The approach of fuzzy physics \cite{thesis,thesis1}  $1)$ to quantum geometry and $2)$ to non-perturbative field theory insists on the  use of finite
 dimensional matrix algebras with suitable Laplacians ( metrics ) to describe the geometry \cite{CONNES}. Fields will be described by the same matrix
  algebras or more precisely by the corresponding projective modules \cite{DENJ2006} and the  action functionals will be given by finite dimensional
   matrix models similar to the IKKT models \cite{IKKT,IKKT1}.

As it turns out a topology change can occur naturally if we  try
to unify spactime and fields using this language of finite
dimensional matrices. This  is precisely the picture which
emerges from the perturbative and non-perturbative studies of
noncommutative  gauge theory on the fuzzy sphere
 \cite{madore,iso}. Indeed  we found in the one-loop calculation \cite{ref} as well as in numerical simulations \cite{subrata,ref11} and  the large $N$
 analysis \cite{ref10} that the noncommutative gauge model on the fuzzy sphere written in \cite{ars} ( which is obtained in the zero-slope limit of string
  theory ) and its  generalizations  undergo  first order phase transitions from the "fuzzy sphere" phase to a "matrix" phase where the fuzzy sphere vacuum
  collapses under quantum fluctuation. The matrix phase is the space consisting of a single point. This topology change from the two dimensional sphere
   to a single point and vice versa is intrinsically a quantum mechanical process and it is related to the perturbative UV-IR mixing phenomena. This result was extended to the case of
    fuzzy ${\bf S}^2\times {\bf S}^2$ in \cite{NCFQED}.

In this article we will go one step further and generalize this
result to higher fuzzy complex projective spaces. In particular
 we will work out the case of fuzzy ${\bf CP}^2$ in detail. We find the possibility of  first order phase transitions between ${\bf CP}^2$ and
  ${\bf S}^2$, between ${\bf CP}^2$ and a matrix phase and between  ${\bf S}^2$ and a matrix phase. This richer structure of topology changes is
   due to the fact that $SU(3)$ contains also $SU(2)$ as a subgroup besides the trivial abelian subgroups $U(1)$. As we will explain generalization of our calculation to higher fuzzy complex projective spaces is obvious and straightforward.

For other approaches to topology change using finite dimensional
matrix algebra and fuzzy physics see  \cite{paulo,hoppe}.

This article is organized as follows.

\tableofcontents

\section{Fuzzy ${\bf CP}^2$}
In this section we will follow \cite{CP2,cp2,brian}. Let $T_a$, $a=1,...,8$, be the generators of $SU(3)$ in the symmetric irreducible
 representation $(n,0)$ of dimension $N=\frac{1}{2}(n+1)(n+2)$. They satisfy
\begin{eqnarray}
[T_a,T_b]=if_{abc}T_c~\label{comm}
\end{eqnarray}
and
\begin{eqnarray}
T_a^2=\frac{1}{3}n(n+3)\equiv
|n|^2~,~d_{abc}T_aT_b=\frac{2n+3}{6}T_c.\label{idd}
\end{eqnarray}
Let $t_a=\frac{{\lambda}_a}{2}$ ( where ${\lambda}_a$ are the usual Gell-Mann matrices )  be the generators of $SU(3)$ in the fundamental
 representation $(1,0)$ of dimension $N=3$. They also satisfy
\begin{eqnarray}
&&2t_at_b=\frac{1}{3}{\delta}_{ab}+(d_{abc}+if_{abc})t_c\nonumber\\
&&tr_3t_at_b=\frac{1}{2}{\delta}_{ab}~,~tr_3t_at_bt_c=\frac{1}{4}(d_{abc}+if_{abc}).
\end{eqnarray}
The $N-$dimensonal generator $T_a$ can be obtained by taking the symmetric product of $n$ copies of the fundamental $3-$dimensional generator $t_a$, viz
\begin{eqnarray}
T_a=(t_a{\otimes}{\bf 1}{\otimes}...{\otimes}{\bf 1}+{\bf 1}{\otimes}t_a{\otimes}...{\otimes}{\bf 1}+...+{\bf 1}{\otimes}{\bf 1}{\otimes}...{\otimes}t_a)_{\rm symmetric}.
\end{eqnarray}
In the continuum ${\bf CP}^2$ is the space of all unit vectors $|\psi>$ in ${\bf C}^3$ modulo the phase. Thus $e^{i\theta}|\psi>$,
for all $\theta {\in}[0,2\pi[$, define the same point on ${\bf CP}^2$. It is obvious that all these vectors $e^{i\theta}|\psi>$ correspond to
 the same projector $P=|\psi><\psi|$. Hence ${\bf CP}^2$ is the space of all projection operators of rank one on ${\bf C}^3$. Let ${\bf H}_N$
  and ${\bf H}_3$ be the Hilbert spaces of the $SU(3)$ representations  $(n,0)$ and $(1,0)$ respectively.
We will define fuzzy ${\bf CP}^2$ through  the canonical $SU(3)$ coherent states as follows. Let $\vec{n}$ be a vector in ${\bf R}^8$, then
 we define the projector
\begin{eqnarray}
P_3=\frac{1}{3}{\bf 1}+n_at_a
\end{eqnarray}
The requirement $P_3^2=P_3$ leads to the condition that $\vec{n}$ is a point on ${\bf CP}^2$ satisfying the equations

\begin{eqnarray}
[n_a,n_b]=0~,~n_a^2=\frac{4}{3}~,~d_{abc}n_an_b=\frac{2}{3}n_c.
\end{eqnarray}
We can write
\begin{eqnarray}
P_3=|\vec{n},3><3,\vec{n}|.
\end{eqnarray}
We think of $|\vec{n},3>$ as the coherent state in ${\bf H}_3$ ( level $3\times 3$ matrices ) which is localized at
the point $\vec{n}$ of  ${\bf CP}^2$. Therefore the coherent state $|\vec{n},N>$ in ${\bf H}_N$ ( level $N\times N$ matrices )
which is localized around the point $\vec{n}$ of  ${\bf CP}^2$ is defined by the projector
\begin{eqnarray}
P_N=|\vec{n},N><N,\vec{n}|=(P_3{\otimes}P_3{\otimes}...{\otimes}P_3)_{\rm symmetric}.
\end{eqnarray}
We compute that
\begin{eqnarray}
tr_3t_aP_3=<\vec{n},3|t_a|\vec{n},3>=\frac{1}{2}n_a~,~
tr_NT_aP_N=<\vec{n},N|T_a|\vec{n},N>=\frac{n}{2}n_a.
\end{eqnarray}
Hence it is natural to identify fuzzy ${\bf CP}^2$ at level $N=\frac{1}{2}(n+1)(n+2)$ ( or  ${\bf CP}^2_n$ ) by the coordinates operators

\begin{eqnarray}
x_a=\frac{2}{n}T_a.
\end{eqnarray}
They satisfy

\begin{eqnarray}
[x_a,x_b]=\frac{2i}{n}f_{abc}x_c~,~x_a^2=\frac{4}{3}(1+\frac{3}{n})~,~d_{abc}x_ax_b=\frac{2}{3}(1+\frac{3}{2n})x_c.
\end{eqnarray}
Therefore in the large $N$ limit we can see that the algebra of $x_a$ reduces to the continuum algebra of $n_a$. Hence $x_a{\longrightarrow}n_a$
 in the continuum limit $N{\longrightarrow}{\infty}$.

The algebra of ~functions on fuzzy  ${\bf CP}^2_n$ is identified with the algebra of $N{\times}N$ matrices $Mat_N$ generated by
 all polynomials in the coordinates operators $x_a$. Recall that $N=\frac{1}{2}(n+1)(n+2)$. The left action of $SU(3)$
  on this algebra is generated by $(n,0)$ whereas the right action is generated by $(0,n)$. Thus the algebra $Mat_N$ decomposes
  under the action of $SU(3)$ as
\begin{eqnarray}
(n,0){\otimes}(0,n)={\otimes}_{p=0}^n(p,p).
\end{eqnarray}
A general function on fuzzy  ${\bf CP}^2_n$ is therefore written as

\begin{eqnarray}
F=\sum_{p=0}^nF_{I^2,I_3,Y}^{(p)}T_{I^2,I_3,Y}^{(p,p)}
\end{eqnarray}
$T_{I^2,I_3,Y}^{(p,p)}$ are $SU(3)$ polarization tensors in the irreducible representation $(p,p)$. $I^2,I_3$ and $Y$ are the square of the isospin,
the third component of the isospin and the hypercharge quantum numbers which characterize $SU(3)$ representations.

The derivations on fuzzy  ${\bf CP}^2_n$ are defined by the commutators $[T_a,..]$. The Laplacian is then obviously given by ${\Delta}_N=[T_a,[T_a,...]]$.
Fuzzy ${\bf CP}^2_n$ is completely determined by the spectral triple ${\bf CP}^2_n=(Mat_N,{\Delta}_N,{\bf H}_N)$. Now we can compute

\begin{eqnarray}
tr_NFP_N=<\vec{n},N|F|\vec{n},N>=F_N(\vec{n})=\sum_{p=0}^nF_{I^2,I_3,Y}^{(p)}Y_{I^2,I_3,Y}^{(p,p)}(\vec{n})
\end{eqnarray}
$Y_{I^2,I_3,Y}^{(p,p)}(\vec{n})$ are $SU(3)$ polarization tensors defined by

\begin{eqnarray}
Y_{I^2,I_3,Y}^{(p,p)}(\vec{n})=<\vec{n},N|T_{I^2,I_3,Y}^{(p,p)}|\vec{n},N>.
\end{eqnarray}
Furthermore we can compute

\begin{eqnarray}
tr_N[T_a,F]P_N=<\vec{n},N|[T_a,F]|\vec{n},N>=({\cal L}_aF_N)(\vec{n})~,~{\cal L}_a=-if_{abc}n_b{\partial}_c.
\end{eqnarray}
And
\begin{eqnarray}
tr_NFGP_N=<\vec{n},N|FG|\vec{n},N>=F_N*G_N(\vec{n}).
\end{eqnarray}
The star product on fuzzy ${\bf CP}^2_n$ is found to be given by \cite{brian}
\begin{eqnarray}
&&F_N*G_N(\vec{n})=\sum_{p=0}^n\frac{(n-p)!}{p!n!}K_{a_1b_1}...K_{a_pb_p}{\partial}_{a_1}...{\partial}_{a_p}F_N(\vec{n}){\partial}_{b_1}...
{\partial}_{b_p}G_N(\vec{n})~\nonumber\\
&&~K_{ab}=\frac{2}{3}{\delta}_{ab}-n_an_b+(d_{abc}+if_{abc})n_c.
\end{eqnarray}
\section{Fuzzy gauge fields on ${\bf CP}^2_n$}
We will introduce fuzzy gauge fields $A_a$, $a=1,...,8$, through the covariant derivatives $D_a$, $a=1,...,8$, as follows
\begin{eqnarray}
D_a=T_a+A_a.
\end{eqnarray}
$D_a$ are $N{\times}N$ matrices which transform under the action
of $U(N)$ as follows $D_a{\longrightarrow}UD_aU^{+}$ where
$U{\in}U(N)$. Hence $A_a$ are   $N{\times}N$ matrices which
transform as $A_a{\longrightarrow}UA_aU^{+}+U[T_a,U^{+}]$. In
order that the field $\vec{A}$ be a $U(1)$ gauge field on fuzzy
${\bf CP}^2_n$ it must satisfies some additional constraints so
that only four of its components are non-zero. These are the
tangent
 components to ${\bf CP}^2_n$. The other four components of $\vec{A}$ are normal to ${\bf CP}^2_n$ and in general they will be projected out from the model.

Let us go back to the continuum ${\bf CP}^2$ and let us consider a gauge field $A_a$\footnote{Remark that we are using the same symbol as in the fuzzy case.
 However this $A_a$ is a function on continuum ${\bf CP}^2$ as opposed to the $A_a$ in the fuzzy setting which is an $N{\times}N$ matrix. }, $a=1,...,8$,
 which is strictly tangent to ${\bf CP}^2$ . By construction this gauge field must satisfy
\begin{eqnarray}
A_a=P_{ab}^TA_b~,~P^T=(n_aAdt_a)^2.\label{cond}
\end{eqnarray}
$P^T$ is the projector which defines the tangent bundle over ${\bf CP}^2$. The normal bundle over ${\bf CP}^2$ will be defined by the projector
$P^N=1-P^T$. Explicitly these are  given by
\begin{eqnarray}
P^T_{ab}=n_cn_d(Adt_c)_{ae}(Adt_d)_{eb}=n_cn_df_{cae}f_{dbe}~,~P^N_{ab}={\delta}_{ab}-n_cn_df_{cae}f_{dbe}.
\end{eqnarray}
In above we have used the fact that the generators in the adjoint representation $(1,1)$ satisfy $(Adt_a)_{bc}=-if_{abc}$.
 Remark that we have the identities $n_aP^T_{ab}=n_bP^T_{ab}=0$. Hence the condition (\ref{cond}) takes the natural form
\begin{eqnarray}
n_aA_a=0.\label{cond1}
\end{eqnarray}
This is one condition which allows us to reduce the number of independent components of $A_a$ by one. We know that there must be three
 more independent constraints which the tangent field $A_a$ must satisfy since it has only $4$ independent components. To find them we
  start from the identity \cite{stein}
\begin{eqnarray}
d_{abk}d_{cdk}=\frac{1}{3}\bigg[{\delta}_{ac}{\delta}_{bd}+{\delta}_{bc}{\delta}_{ad}-{\delta}_{ab}{\delta}_{cd}+f_{cak}f_{dbk}+f_{dak}f_{cbk}\bigg].\label{24}
\end{eqnarray}
Thus
\begin{eqnarray}
n_cn_dd_{abk}d_{cdk}=\frac{2}{3}\bigg[{n}_{a}{n}_{b}-\frac{2}{3}{\delta}_{ab}+n_cn_df_{cak}f_{dbk}\bigg].
\end{eqnarray}
By using the fact that $d_{cdk}n_cn_d=\frac{2}{3}n_k$ we obtain

\begin{eqnarray}
d_{abk}n_k={n}_{a}{n}_{b}-\frac{2}{3}{\delta}_{ab}+n_cn_df_{cak}f_{dbk}.\label{plk55}
\end{eqnarray}
Hence it is a straightforward calculation to find that the gauge
field $A_a$ must also satisfy the conditions
\begin{eqnarray}
d_{abk}n_kA_b=\frac{1}{3}A_a.\label{cond2}
\end{eqnarray}
In the case of ${\bf S}^2$ the projector $P^T$ takes the simpler form $P^T_{ab}={\delta}_{ab}-n_an_b$ and hence $P^N_{ab}=n_an_b$.
 From equation (\ref{plk55}) we have on ${\bf CP}^2$
\begin{eqnarray}
P_{ab}^T=d_{abc}n_c-n_an_b+\frac{2}{3}{\delta}_{ab}~,~P_{ab}^N=-d_{abc}n_c+n_an_b+\frac{1}{3}{\delta}_{ab}.
\end{eqnarray}
If we choose to sit on the ``north pole'' of ${\bf CP}^2$, i.e $\vec{n}=(0,0,0,0,0,0,0,-\frac{2}{\sqrt{3}})$
then we can find that $P^T=diag(0,0,0,1,1,1,1,0)$ and as a consequence $P^N=(1,1,1,0,0,0,0,1)$ . So $Adt_a$, $a=1,2,3,8$ correspond to
 the normal directions while $Adt_a$, $a=4,5,6,7$ correspond to the tangent directions.

Indeed by substituting $\vec{n}=(0,0,0,0,0,0,0,-\frac{2}{\sqrt{3}})$ in equation (\ref{cond2})
 and using $d_{8ij}=\frac{1}{\sqrt{3}}{\delta}_{ij}$ where $i,j=1,2,3$ and $d_{8\alpha \alpha}=-\frac{1}{2\sqrt{3}}$
 where $\alpha =4,5,6,7$ and $d_{888}=-\frac{1}{\sqrt{3}}$ we get $A_1=A_2=A_3=A_8=0$ which is what we want. In fact (\ref{cond2})
 already contains (\ref{cond1}). In other words it contains exactly the correct number of equations needed to project out the gauge field
 $A_a$ onto the tangent bundle of ${\bf CP}^2$.

Let us finally say that given any continuum gauge field $A_a$ which does not satisfy the constraints (\ref{cond1}) and (\ref{cond2})
 we can always make it tangent by applying the projector $P^T$. Thus we will have the tangent gauge field
\begin{eqnarray}
A_a^T=P_{ab}^TA_b=d_{abc}n_cA_b-n_a(n_bA_b)+\frac{2}{3}A_a.
\end{eqnarray}
Similarly the fuzzy gauge field must satisfy some conditions  which should reduce to (\ref{cond1}) and (\ref{cond2}).
As it turns out constructing a tangent fuzzy gauge field using an expression like (\ref{cond}) is a highly non-trivial task due to   $1)$
 gauge covariance problems and  $2)$ operator ordering problems. However implementing (\ref{cond1}) and (\ref{cond2}) in the fuzzy setting
 is quite easy  since we will only need to return to the covariant derivatives $D_a$ and require them to satisfy the $SU(3)$ identities  (\ref{idd}), viz
\begin{eqnarray}
&&D_a^2=\frac{1}{3}n(n+3)\nonumber\\
&&d_{abc}D_aD_b=\frac{2n+3}{6}D_c.\label{cond3}
\end{eqnarray}
So $D_a$ are almost the $SU(3)$ generators except that they fail to satisfy the  fundamental commutation relations of $SU(3)$ given by equation (\ref{comm}).
This failure is precisely measured by the curvature of the gauge field $A_a$, namely
\begin{eqnarray}
F_{ab}&=&i[D_a,D_b]+f_{abc}D_c\nonumber\\
&=&i[T_a,A_b]-i[T_b,A_a]+f_{abc}A_c+i[A_a,A_b].
\end{eqnarray}
The continuum limit of this object is clearly given by the usual
curvature on ${\bf CP}^2$, viz $F_{ab}=i{\cal L}_aA_b-i{\cal
L}_bA_a+f_{abc}A_c+i[A_a,A_b]$. To check that this fuzzy gauge
field $A_a$ has the correct degrees of freedom we need to check
that the identities (\ref{cond3}) reduce to (\ref{cond1}) and
(\ref{cond2}) in the continuum limit $n{\longrightarrow}\infty $.
This fact is quite straightforward to verify and we leave it as an
exercise.

Next we need to write down actions on fuzzy ${\bf CP}^2_n$. The first piece is the usual Yang-Mills action
\begin{eqnarray}
S_{\rm YM}=\frac{1}{4g^2}Tr_NF_{ab}^2.
\end{eqnarray}
By construction it has the correct continuum limit. $Tr_N$ is the normalized trace $Tr_N{\bf 1}=1$.

The second  piece in the action is a potential term which has to
implement the constraints (\ref{cond3}) in some limit. Indeed we
will not impose these constraints rigidly on the path integral but
we will include their effects by adding to the action a very
special potential term. In other words we will not assume that
$D_a$ satisfy (\ref{cond3}). To the end of writing this potential
term  we will  introduce the four normal scalar fields on fuzzy
${\bf CP}^2_n$ by the equations ( see equations (\ref{cond3}) )

\begin{eqnarray}
&&{\Phi}=\frac{1}{n}(D_a^2-\frac{1}{3}n(n+3))=\frac{1}{2}x_aA_a+\frac{1}{2}A_ax_a+\frac{1}{n}A_a^2{\longrightarrow}n_aA_a
\end{eqnarray}
and
\begin{eqnarray}
{\Phi}_c=\frac{1}{n}(d_{abc}D_aD_b-\frac{2n+3}{6}D_c)&=&\frac{1}{2}d_{abc}x_aA_b+\frac{1}{2}d_{abc}A_ax_b-\frac{2n+3}{6n}A_c+\frac{1}{n}d_{abc}A_aA_b\nonumber\\
&{\longrightarrow}&d_{abc}n_aA_b-\frac{1}{3}A_c.\label{cond4}
\end{eqnarray}
We add to the Yang-Mills action the potential term
\begin{eqnarray}
V_0=\beta Tr_N{\Phi}^2 +M^2 Tr_N{\Phi}_a^2.
\end{eqnarray}
In the limit where the parameters $\beta$ and $M^2$ are taken to be very large positive numbers we can see that only configurations $A_a$
 ( or equivalently $D_a$ ) such that $\Phi=0$ and ${\Phi}_c=0$ dominate the path integral which is precisely what we want. This is the
  region of the phase space of most interest. This is the classical prediction.

However in the quantum theory we will find that the parameter $\beta$ must be related to $M^2$ in
some specific way  in order to kill  exactly the normal components of $A_a$. This result ( which we will show shortly
in the one-loop quantum fuzzy theory  ) is the quantum analogue of  the classical continuum statement that equation (\ref{cond2})
contains already (\ref{cond1}).

\section{The classical and one-loop quantum actions on ${\bf CP}^2_n$}
The total action is then given by
\begin{eqnarray}
S_1&=&\frac{1}{2g^2}Tr_NF_{ab}^2+\beta Tr_N{\Phi}^2+M^2 Tr_N{\Phi}_c^2\nonumber\\
&=&\frac{1}{g^2}Tr_N\bigg[-\frac{1}{4}[D_a,D_b]^2+if_{abc}D_aD_bD_c\bigg]+\frac{3n}{4g^2}Tr_N\Phi +\beta Tr_N{\Phi}^2+M^2 Tr_N{\Phi}_c^2.
\end{eqnarray}
This is essentially the same action  considered
 in \cite{stein}.
However this action is  different from the action considered in \cite{subrata} which is of the form
\begin{eqnarray}
  S_0&=&\frac{1}{g^2}Tr_N\bigg[-\frac{1}{4}[D_a,D_b]^2+\frac{2i}{3}f_{abc}D_aD_bD_c\bigg].
\end{eqnarray}
The first difference is between the cubic terms which come with different coefficients. The second more crucial difference is
 the presence of the potential term in our case. The linear term in $\Phi$ is actually a part of the Yang-Mills action.

The equations of motion derived from the action $S_0$ are
\begin{eqnarray}
[D_a,F_{ab}]=0.
\end{eqnarray}
These are solved by the fuzzy ${\bf CP}^2_n$ configurations
\begin{eqnarray}
D_a=T_a \label{cp2nf}
\end{eqnarray}
and also by the diagonal matrices
\begin{eqnarray}
D_a={\rm diag}(d_a^1,d_a^2,...,d_a^N). \label{cp2di}
\end{eqnarray}
We think of these diagonal matrices ( including the zero matrix ) as describing a single point in accordance with the IKKT model \cite{IKKT}.

More interestingly is the fact that these equations of motion are also solved by the fuzzy ${\bf S}^2_N$
 configurations 
\begin{eqnarray}
D_i=L_i~,~i=1,2,3~{\rm and}~D_{\alpha}=0~{\alpha}=4,5,6,7,8.\label{cp2s2}
\end{eqnarray}
Indeed in this case $[D_{a},F_{ab}]=[D_i,F_{ib}]$. For $b=j$ this
is equal to $0$ because $f_{ijk}={\epsilon}_{ijk}$ whereas for
$b=\alpha$ this is equal to zero because $f_{ij\alpha}=0$. In
above $L_i$ are the generators of $SU(2)$ in the irreducible
representation $\frac{N-1}{2}$.

The equations of motion derived from the action $S_1$  are on the other hand given by
\begin{eqnarray}
\frac{i}{g^2}[D_b,F_{ab}]+\frac{1}{2g^2}f_{abc}F_{bc}+2\beta\{\Phi,D_a\}+M^2\bigg(2d_{abc}\{{\Phi}_c,D_b\}-\frac{2n+3}{3}{\Phi}_a\bigg)=0.
\end{eqnarray}
Now the only solutions of these equations of motion are the ${\bf CP}^2_n$ configurations (\ref{cp2nf}). Thus the potential term has eliminated the
 diagonal matrices (\ref{cp2di}) as possible solutions. In fact this classical observation will not hold in the quantum theory for all values of
 the parameter  $M^2$ since there will always be a region in the phase space of the theory where the vacuum solution is not $D_a=T_a$ but $D_a=0$.
  However when we take $M^2$ to be very large positive number then we can see that $D_a=T_a$ becomes quantum mechanically more stable. Hence by
  neglecting the potential term we can not at all speak of the space ${\bf CP}^2_n$ since it will collapse rather quickly under quantum fluctuations
  to a single point.

The potential term has also eliminated the fuzzy ${\bf S}^2_N$
configurations (\ref{cp2s2}) as possible solutions. In fact even if we set $M=\beta=0$ in the above equations of motion the fuzzy ${\bf S}^2_N$
configurations (\ref{cp2s2}) are not solutions.

The other major difference between $S_1$ and $S_0$ is that if we expand around the fuzzy ${\bf CP}^2_n$ solution $D_a=T_a$ by writing $D_a=T_a+A_a$ and
 then substitute back in $S_1$ and $S_0$ we find that $S_0$ does not yield in the continuum limit the usual pure gauge theory on ${\bf CP}^2$.
 It contains an extra piece which resembles the Chern-Simons action ( although it is strictly real ). We skip here the corresponding elementary proof
 ( see the appendix ). $S_1$ will yield on the other hand the desired pure gauge theory on ${\bf CP}^2_n$ in the limit $M^2 {\longrightarrow}\infty$ and
  hence it has the correct continuum limit. If we do not take the limit $M^2{\longrightarrow}\infty$ then $S_1$ will describe a gauge theory coupled to $4$
  adjoint scalar fields  which are the  normal components of $A_a$.

The only motivation for $S_0$-as far as we can see- is its similarity to the fuzzy ${\bf S}^2$ action which looks precisely like $S_0$ with
the replacement $f_{abc}{\longrightarrow}{\epsilon}_{abc}$. This fuzzy sphere action was obtained in string theory in
the limit ${\alpha}^{'}{\longrightarrow}0$ when we have open strings moving in a curved background with an ${\bf S}^3$ metric in
 the presence of a Neveu-Schwarz B-field.
However it is found that perturbation theory with $S_0$ is simpler than perturbation theory with $S_1$. More importantly it is found that $S_0$
allows for some new topology change which does not occur with $S_1$. In particular the transitions ${\bf CP}^2_n{\longrightarrow}{\bf S}^2_N$ and
 ${\bf S}^2_N{\longrightarrow}{\bf CP}^2_n$ are  possible in the quantum theory of $S_0$.

Using the background field method we find that the one-loop effective action in the gauge ${\xi}^{-1} =1+\frac{2g^2{\beta}_i}{n^2}$ is given by
( see the appendix  )

\begin{eqnarray}
{\Gamma}_i[D]&=&{S}_i[D]+\frac{1}{2}Tr_8TR\log{\Omega}^i_{ab}-TR\log{\cal D}_a^2.\label{for}
\end{eqnarray}
Where
\begin{eqnarray}
{S}_i[D]&=&\frac{1}{g^2}Tr_N\bigg(-\frac{1}{4}[D_a,D_b]^2+i{\alpha}_i f_{abc}D_aD_bD_c\bigg)+{\rho}_i Tr_N\Phi +{\beta}_iTr_N{\Phi}^2+M_i^2 Tr_N{\Phi}_c^2.\nonumber\\
\end{eqnarray}
And
\begin{eqnarray}
{\Omega}^i_{ab}&=&{\cal D}_c^2{\delta}_{ab}-2i{\cal F}_{ab}+2i(1-\frac{3{\alpha}_i}{2})f_{abc}{\cal D}_c+\frac{2g^2{\rho}_i}{n}{\delta}_{ab}+\frac{2g^2{\beta}_i}{n^2}\big(4D_{a}D_{b}+2n{\delta}_{ab}\Phi\big)+\frac{2g^2M_i^2}{n^2}\bigg(\nonumber\\
&-&d_{aa^{'}c}d_{bb^{'}c}{\cal D}_{a^{'}}{\cal D}_{b^{'}}
+4d_{aa^{'}c}d_{bb^{'}c}D_{a^{'}}D_{b^{'}}-2(\frac{2n+3}{3})d_{abc}D_c+2nd_{abc}{\Phi}_c
+(\frac{2n+3}{6})^2{\delta}_{ab}\bigg).\nonumber\\
\end{eqnarray}
The trace $TR$ corresponds to the left and right actions of
operators on matrices whereas $Tr_8$ is the trace associated with
$8-$dimensional rotations. Given a matrix  $O$ the operator ${\cal
O}$ is given by ${\cal O}(..)=[O,..]$, for example ${\cal
D}_a(Q_a)=[D_a,Q_a]$. For $S_0$ we have ${\alpha}_0=\frac{2}{3}$,
${\rho}_0={\beta}_0=M_0=0$ while for $S_1$  we have
${\alpha}_1=1$,  ${\rho}_1 =\frac{3n}{4g^2}$, ${\beta}_1=\beta$
and $M_1=M$.

\section{Fuzzy ${\bf CP}^2$  phase and a stable fuzzy sphere phase}

\paragraph{Fuzzy ${\bf CP}^2$ phase}

Let us first neglect the potential term in $S_i$, i.e we will set ${\beta}_i=M_i=0$ or equivalently $V_0=0$ in $S_1$. The effective potential
is given by the formula (\ref{for}), viz
\begin{eqnarray}
{\Gamma}_i[D]&=&S_i[D]+\frac{1}{2}Tr_8TR\log{\Omega}^i_{ab}-TR\log{\cal D}_a^2
\end{eqnarray}
where the background field is chosen such that
\begin{eqnarray}
D_a=\phi T_a.\label{cp2-conf}
\end{eqnarray}
The reason is simply because we want to study the stability of the
fuzzy ${\bf CP}^2_n$ vacuum $D_a=T_a$ against quantum
fluctuations. Hence $\phi$ is an order parameter which measures in
a well defined obvious sense the radius of  ${\bf CP}^2_n$. Let us
compute the
 classical potential in this configuration. we have
\begin{eqnarray}
S_i[D]
&=&\frac{3|n|^2}{g^2}\bigg[\frac{1}{4}{\phi}^4-\frac{{\alpha}_i}{2}{\phi}^3+\frac{g^2{\rho}_i}{3n}({\phi}^2-1)\bigg].\label{clcp2}
\end{eqnarray}
The main quantum correction is equal to the trace of the logarithm
of the Laplacian ${\Omega}^i$ which is given by the simple formula
\begin{eqnarray}
{\Omega}^i_{ab}&=&\bigg({\phi}^2{\cal L}_c^2+\frac{2g^2{\rho}_i}{n}\bigg){\delta}_{ab}+2i\phi(\phi -\frac{3{\alpha}_i}{2})f_{abc}{\cal L}_c\label{plk}
\end{eqnarray}
where ${\cal L}_a(..)=[T_a,..]$. There are two cases to consider. For both $S_0$ and $S_1$ we obtain the quantum correction ( see the appendix  )

\begin{eqnarray}
{\Delta}{\Gamma}_i[D]&=&\frac{1}{2}Tr_8TR\log({\phi}^2{\cal L}_a^2{\bf 1}_8)-TR\log({\phi}^2{\cal L}_a^2)\nonumber\\
&=&+6N^2\log{\phi}+{\rm constant}.
\end{eqnarray}

\paragraph{Case $1$} For $S_0$ we have ${\alpha}_0 =\frac{2}{3}$ and ${\rho}_0 =0$.
Thus the quantum effective potential is
\begin{eqnarray}
V_{\rm eff}=\frac{{\Gamma}_0[D]}{6N^2}
&=&\frac{2}{3n^2g^2}\bigg[\frac{1}{4}{\phi}^4-\frac{1}{3}{\phi}^3\bigg]+\log{\phi}+{\rm constant}.\label{rmconstant}
\end{eqnarray}
The quantum minimum of the model is given by the value of $\phi$ which solves the equation $V_{\rm eff}^{'}=0$. It is not difficult
to convince ourselves that this equation of motion will admit a solution only up to an upper critical value $g_{*}$ of the gauge
coupling constant $g$ beyond which the configuration $D_a=\phi T_a$ collapses. At this value $g_{*}$ the potential $V_{\rm eff}$
becomes unbounded from below. The conditions which will yield the critical value $g_{*}$ are therefore
\begin{eqnarray}
&&V_{\rm eff}^{'}=\frac{2}{3n^2g^2}\bigg[{\phi}^3-{\phi}^2\bigg]+\frac{1}{\phi}=0~,~
V_{\rm eff}^{''}=\frac{2}{3n^2g^2}\bigg[3{\phi}^2-2{\phi}\bigg]-\frac{1}{{\phi}^2}=0.\label{firs}
\end{eqnarray}
We find immediately
\begin{eqnarray}
{\phi}_{*}=\frac{3}{4}~,~n^2g_{*}^2=\frac{2}{9}(\frac{3}{4})^4=0.0703.\label{crit}
\end{eqnarray}
Above the value $g_{*}$ we do not have a fuzzy ${\bf CP}^2_n$, in other words the space ${\bf CP}^2_n$ evaporates at this point.
This critical point separates two distinct phases of the model, in the region above $g_*$ we have a  ``matrix phase'' while in the
 region below $g_{*}$ we have a ``fuzzy ${\bf CP}^2_n$'' phase in which the model admits the interpretation of being a $U(1)$ gauge
 theory on ${\bf CP}^2$.

By going from small values of $g$ ( $g{\leq}g_{*}$  corresponding to the ``fuzzy ${\bf CP}_n^2$ phase'' ) towards large values of $g$ we get through
 the value $g_{*}$ where the space ${\bf CP}^2_n$ decays.
Looking at this process the other way around we can see that starting from large values of $g$ ( $g{>}g_{*}$  corresponding to the ``matrix phase'' )
 and going through $g_*$ we generate the space ${\bf CP}^2_n$ dynamically. It seems therefore that we have generated quantum mechanically the spectral
  triple which defines the space ${\bf CP}^2_n$.

Furthermore we note that in the ``matrix phase'' we have a  $U(N)$ gauge theory reduced to a point where $N$ is the size of the matrices since
 the minimum there is given by diagonal matrices. The important point is that in this phase the gauge group is certainly not $U(1)$.
 Let us  recall that in the ``fuzzy ${\bf CP}^2_n$ phase'' we had a $U(1)$ gauge theory. Hence across the transition line between the
  ``fuzzy ${\bf CP}^2_n$ phase'' and the ``matrix phase'' the structure of the gauge group also changes. Thus we obtain in this model in
  correlation with the topology change across the critical line a novel spontaneous symmetry breaking mechanism.

For the purpose of comparing with the numerical results of ~\cite{subrata} we define the coupling constant
$\bar{\alpha}$ such that $\bar{\alpha}^4N=\frac{1}{g^2}$. Then the critical value (\ref{crit}) is seen to occur at
\begin{eqnarray}
\bar{\alpha}_{*}=2.309.
\end{eqnarray}
This is precisely the result of the Monte Carlo simulation reported in equation $(3.2)$ of \cite{subrata}.
\paragraph{Case $2$} For $S_1$ we have ${\alpha}_1 =1$ and ${\rho}_1 =\frac{3n}{4g^2}$ and hence the effective potential is given by

\begin{eqnarray}
V_{\rm eff}=\frac{{\Gamma}_1[D]}{6N^2}
&=&\frac{2}{3n^2g^2}\bigg[\frac{1}{4}{\phi}^4-\frac{1}{2}{\phi}^3+\frac{1}{4}{\phi}^2\bigg]+\log{\phi}+{\rm constant}.\label{rmconstant1}
\end{eqnarray}
A direct calculation yields the critical values

\begin{eqnarray}
{\phi}_{*}=\frac{9+\sqrt{17}}{16}~,~n^2g_{*}^2=\frac{{\phi}_*^2}{4}({\phi}_* -\frac{2}{3})=0.02552.\label{crit1}
\end{eqnarray}
This $g_{*}$ is smaller than the $g_*$  obtained in (\ref{crit})
and hence the fuzzy ${\bf CP}^2_n$ is more stable in the model
$S_0$ than it is in the model $S_1$ which is largely due to the
linear term proportional to $\Phi$ in $S_1$. In other words
attempting to put true gauge theory on fuzzy ${\bf CP}^2_n$ causes
the space to decay more rapidly. However for $S_0$ the true vacuum
is the fuzzy sphere ${\bf S}^2_N$ and not the fuzzy ${\bf CP}^2_n$
as we will now discuss

\paragraph{A stable fuzzy sphere phase}

We know that there is also a fuzzy sphere solution (\ref{cp2s2}) for the model $S_0$. We consider then the background field
\begin{eqnarray}
D_i=\phi T_i~,~i=1,2,3~,D_{\alpha}=0~,~\alpha =4,5,6,7,8.\label{lmnb}
\end{eqnarray}
We want now to study the stability of this vacuum against quantum
fluctuations. The $\phi$ is now an order parameter which measures
the radius of  the fuzzy sphere ${\bf S}^2_N$. The classical
potential in this configuration is
\begin{eqnarray}
S_0[D]
&=&\frac{2c_2}{g^2}\bigg[\frac{1}{4}{\phi}^4-\frac{1}{3}{\phi}^3\bigg].\label{clcp2s2}
\end{eqnarray}
In above $c_2=\frac{N^2-1}{4}$ is the Casimir of $SU(2)$ in the irreducible representation $\frac{N-1}{2}$ ( $N=\frac{1}{2}(n+1)(n+2)$ ).
It is clear that $2c_2>>3|n|^2$ and $\frac{|n|^2}{c_2}<<1$ in the large $n$ limit.  Hence the action (\ref{clcp2s2}) around the
classical minimum $\phi=1$ is much smaller than the classical action (\ref{clcp2}). In other words the fuzzy sphere is more stable than
 the fuzzy ${\bf CP}^2_n$ in this case.

The quantum corrections are given in this case by
\begin{eqnarray}
\frac{1}{2}Tr_3TR\log {\Omega}_{ij}+\frac{1}{2}Tr_5TR\log\tilde{\Omega}_{\alpha \beta}-TR\log{\phi}^2{\cal L}_i^2.
\end{eqnarray}
In above
\begin{eqnarray}
{\Omega}_{ij}&=&\bigg({\phi}^2{\cal L}_k^2+\frac{2g^2\rho}{n}\bigg){\delta}_{ij}+2i\phi(\phi -\frac{3\alpha}{2}){\epsilon}_{ijk}{\cal L}_k~,~
\tilde{\Omega}_{\alpha \beta}=\bigg({\phi}^2{\cal L}_k^2+\frac{2g^2\rho}{n}\bigg){\delta}_{\alpha \beta}-3i\alpha \phi f_{\alpha \beta i}{\cal L}_i.\nonumber\\
\end{eqnarray}
Following the same arguments of the previous section ( only now it is $SU(2)$ representation theory which is involved ) we have in the large $n$ limit
\begin{eqnarray}
\frac{1}{2}Tr_3TR\log {\Omega}_{ij}-TR\log{\phi}^2{\cal L}_i^2=\frac{1}{2}Tr_3TR\log {\phi}^2{\delta}_{ij}-TR\log{\phi}^2+..=N^2\log\phi +...\label{enarray1}
\end{eqnarray}
We can also argue that we  have
\begin{eqnarray}
\frac{1}{2}Tr_5TR\log {\Omega}_{\alpha \beta}+..=5N^2\log\phi +...\label{enarray2}
\end{eqnarray}
 In other words the configurations (\ref{lmnb}) although they are fuzzy sphere configurations  they know ( through their quantum interactions )
 about the other $SU(3)$ structure present in the model. Classically this $SU(3)$ structure is not detected at all by these configurations in
 the classical potential (\ref{clcp2s2}). The effective potential  becomes in this case

\begin{eqnarray}
V_{\rm eff}=\frac{{\Gamma}_0[D]}{6N^2}
&=&\frac{1}{12g^2}\bigg[\frac{1}{4}{\phi}^4-\frac{1}{3}{\phi}^3\bigg]+\log{\phi}+{\rm constant}.\label{rmconstant2}
\end{eqnarray}
A direct calculation yields the critical value

\begin{eqnarray}
{\phi}_*=\frac{3}{4}~,~g_{*}^2=\frac{1}{36}(\frac{3}{4})^4=0.0087875.\label{763}
\end{eqnarray}
In terms of the coupling $\tilde{\alpha}$ define by $\tilde{\alpha}^4=\frac{1}{g^2}$ the critical value $g_*$ reads
\begin{eqnarray}
\tilde{\alpha}_*=3.26.
\end{eqnarray}
This is again what is measured in the Monte Carlo simulation of the model $S_0$ as it is reported in equation $(4.2)$ of \cite{subrata}.
Therefore we have a fuzzy sphere phase above  $\tilde{\alpha}_{*}$ and a matrix phase below $\tilde{\alpha}_{*}$. The model $S_0$ can also be
in a fuzzy ${\bf CP}^2_n$ phase for $n^2g_*^2$ below the second  value of (\ref{crit}) which for large enough $n$ is much smaller than the value
  $n^2g_*^2$ with $g_*^2$  given by the second equation of (\ref{763}). However we have seen in the previous paragraph that this ${\bf CP}_n^2$ will
   decay rather quickly to a single point which ( by the  discussion of the present section ) can only happen  by going first across a fuzzy sphere phase.
    We have then the transition pattern  ${\bf CP}^2_n{\longrightarrow}{\bf S}^2_N{\longrightarrow}\{0\}$.
In the limit where $\bar{\alpha}^4={2}/{n^2g^2}$ is kept fixed we can see that the above critical value (\ref{763}) is infinitely large
 which means that the model $S_0$ is mostly in the fuzzy sphere phase. The matrix phase shrinks to zero and the fuzzy sphere is completely
  stable in this limit since  the  fuzzy ${\bf CP}^2_n$ phase can occur only at  very small values of the coupling constant $n^2g^2$.

\section{The large mass limit and the transition ${\bf CP}^2{\longrightarrow}{\bf S}^2$}
\paragraph{The large mass limit}

 Now we include the effect of the potential term $V_0$. The relevant model is given by the action $S_1$. Naturally the calculation becomes
  more complicated in this case. The classical  potential in the configuration $D_a=\phi T_a$ is

\begin{eqnarray}
 \frac{S_1[D]}{6N^2}
&=&\frac{2}{3n^2g^2}\bigg[\frac{1}{4}{\phi}^4-\frac{1}{2}{\phi}^3+\frac{1}{4}{\phi}^2+\frac{g^2\beta}{9}({\phi}^2-1)^2+\frac{g^2M^2}{27}({\phi}^2-\phi)^2\bigg].\label{cl}
\end{eqnarray}
The most important quantum correction is given by the determinant of
\begin{eqnarray}
{\Omega}^1_{ab}&=&\bigg({\phi}^2{\cal L}_c^2+\frac{3}{2}\bigg){\delta}_{ab}-2\phi(\phi -\frac{3}{2})(Adt_c{\cal L}_c)_{ab}+2g^2{\beta}{\Delta}_1{\Omega}_{ab}+2g^2M^2{\Delta}_2{\Omega}_{ab}.
\end{eqnarray}
By using the identities (\ref{idd}) and (\ref{24}) we find that the extra contributions are given explicitly by the expressions
\begin{eqnarray}
{\Delta}_1{\Omega}_{ab}=\frac{1}{n^2}\bigg[4{\phi}^2T_aT_b+2({\phi}^2-1)T_c^2{\delta}_{ab}\bigg].
\end{eqnarray}
\begin{eqnarray}
{\Delta}_2{\Omega}_{ab}&=&\frac{1}{n^2}\bigg[(\frac{2n+3}{6})^2{\delta}_{ab}+({\phi}^2
-3{\phi})(\frac{2n+3}{3})d_{abc}T_c\nonumber\\
&+&\frac{4{\phi}^2}{3}\bigg(T_c^2+\frac{1}{16}-(Adt_cT_c-\frac{1}{4})^2\bigg)_{ba}-\frac{{\phi}^2}{3}\bigg({\cal L}_c^2+\frac{1}{16}-(Adt_c{\cal L}_c-\frac{1}{4})^2\bigg)_{ba}\bigg].\label{plkmass}
\end{eqnarray}
In the continuum large $N$ limit the first extra correction behaves as
\begin{eqnarray}
{\Delta}_1{\Omega}_{ab}={\phi}^2n_an_b+\frac{2}{3}({\phi}^2-1){\delta}_{ab}.
\end{eqnarray}
Let us introduce the projector  $\hat{P}_{ab}=\frac{3}{4}n_an_b$. This is a rank one normal projector which projects vector fields along
the normal direction $Adt_8$. Recall the rank four tangent projector $P^T_{ab}=d_{abc}n_c-n_an_b+\frac{2}{3}{\delta}_{ab}$ and the rank four
normal projector $P^N_{ab}=-d_{abc}n_c+n_an_b+\frac{1}{3}{\delta}_{ab}$. Then we must necessarily have $P^N_{ab}=\hat{P}_{ab}+\tilde{P}_{ab}$
where $\tilde{P}_{ab}$ is a rank three normal projector which projects vector fields along the normal directions $Adt_i$, $i=1,2,3$.
It is given by $\tilde{P}_{ab}=-d_{abc}n_c+\frac{1}{4}n_an_b+\frac{1}{3}{\delta}_{ab}$. We have the decomposition $1=P^T+P^N=P^T+\hat{P}+\tilde{P}$.
 Hence
\begin{eqnarray}
{\Delta}_1{\Omega}_{ab}&=&\frac{4}{3}{\phi}^2\hat{P}_{ab}+\frac{2}{3}({\phi}^2-1){\delta}_{ab}\nonumber\\
&=&\frac{2}{3}({\phi}^2-1)P^T_{ab}+2({\phi}^2-\frac{1}{3})\hat{P}_{ab}+\frac{2}{3}({\phi}^2-1)\tilde{P}_{ab}.\label{input0}
\end{eqnarray}
Similarly in the continuum large $N$ limit the first three terms of ${\Delta}_2{\Omega}_{ab}$ takes the form
( by using also the identity $d_{abc}n_c=\frac{1}{3}P_{ab}^T-\frac{2}{3}\tilde{P}_{ab}+\frac{2}{3}\hat{P}_{ab}$ )
\begin{eqnarray}
{\rm First~3~terms~of~}~{\Delta}_2{\Omega}_{ab}&=&-\frac{{\phi}^2}{3}P^T_{ab}+\frac{1+4{\phi}^2}{9}{\delta}_{ab}+\frac{1}{3}({\phi}^2-3\phi)d_{abc}n_c\nonumber\\
&=&\frac{1+2{\phi}^2-3\phi}{9}P_{ab}^T+\frac{1+6{\phi}^2-6\phi}{9}\hat{P}_{ab}+\frac{1+2{\phi}^2+6\phi}{9}\tilde{P}_{ab}.
\label{input}
\end{eqnarray}
Let us remark  that the coefficients in front of the projectors $\hat{P}$ and $\tilde{P}$ are the masses of the normal components of the gauge
 field and hence they must be  positive. For example the mass of the normal
  components $\tilde{P}_{ab}A_b$ is given by $M^2\tilde{m}^2=\frac{2g^2M^2}{9}( 2(1+3\gamma){\phi}^2+6\phi +1-6\gamma)$ where $\gamma=\frac{\beta}{M^2}$.
   This is positive definite for all values $\phi {\geq}0$ of the radius of ${\bf CP}^2_n$ if $\gamma$ is such that
    $1+3\gamma {\geq}0$ and $1-6\gamma {>}0$. Thus $\gamma$ must be  in the range $-\frac{1}{3}{\leq}\gamma{<}\frac{1}{6}$.
     Since $\beta$ must be positive we obtain the condition
\begin{eqnarray}
0{\leq}\beta {<}\frac{M^2}{6}.\label{range0}
\end{eqnarray}
The mass of the normal component $\hat{P}_ {ab}A_b$ is given by  $M^2\hat{m}^2=\frac{2g^2M^2}{9}( 6(1+3\gamma){\phi}^2-6\phi +1-6\gamma)$.
 The requirement that this mass must be positive definite gives now the condition that the radius $\phi$ can only be in the range
\begin{eqnarray}
\phi{<}{\phi}_-\equiv\frac{1-\sqrt{1-12(\frac{1}{3}+\gamma)(\frac{1}{6}-\gamma)}}{6(\frac{1}{3}+\gamma)}~,~{\phi}{>}{\phi}_+\equiv\frac{1+\sqrt{1-12(\frac{1}{3}+\gamma)(\frac{1}{6}-\gamma)}}{6(\frac{1}{3}+\gamma)}.\label{range}
\end{eqnarray}
We remark that for all allowed values of $\gamma$ we have ${\phi}_-{>}0$ and ${\phi}_{+}{<}1$ so we can still access the limits ${\phi}{\longrightarrow}1$ and ${\phi}{\longrightarrow}0_+$ although there is now a forbidden gap between these two important regions.

The mass of the tangent components $P_{ab}^TA_b$ is given by $M^2m_T^2=\frac{4g^2M^2(1+3\gamma )}{9}(\phi -1)(\phi -\frac{1-6\gamma}{2+6\gamma})$.
This is not always positive in the range (\ref{range}). However this mass formally vanishes in the limit  $M{\longrightarrow}\infty$
where the most probable value of the radius of ${\bf CP}^2_n$ is $\phi {\sim}1$. Finally the last correction of the inverse propagator
 coming from the addition of the potential $V_0$ ( which is given explicitly by the last term in (\ref{plkmass})  ) is also negative.
  Remark that this correction is proportional to ${\phi}^2$ and as a consequence we will not need to compute it explicitly ( see below ).

We are now ready  to compute the determinant. We have
\begin{eqnarray}
{\Omega}^1_{ab}={\Omega}_{ab}+M^2m_T^2P_{ab}^T+M^2\hat{m}^2\hat{P}_{ab}+M^2\tilde{m}^2\tilde{P}_{ab}
\end{eqnarray}
where
\begin{eqnarray}
{\Omega}_{ab}&=&\bigg({\phi}^2{\cal L}_c^2+\frac{3}{2}\bigg){\delta}_{ab}-2\phi(\phi -\frac{3}{2})(Adt_c{\cal L}_c)_{ab}-\frac{2g^2M^2{\phi}^2}{3n^2}\bigg({\cal L}_c^2+\frac{1}{16}-(Adt_c{\cal L}_c-\frac{1}{4})^2\bigg)_{ba}.\nonumber\\
\end{eqnarray}
Thus
\begin{eqnarray}
\frac{1}{2}Tr_8TR\log{\Omega}^1
&=&-\log\int dA_a \exp\bigg[-Tr_NA_a{\Omega}_{ab}A_b\nonumber\\
&-&M^2{m}_T^2Tr_N(A_a^T)^2-M^2\hat{m}^2Tr_N(\hat{A}_a)^2-M^2\tilde{m}^2Tr_N(\tilde{A}_a)^2\bigg].
\end{eqnarray}
From the last two terms we get in the large $M$ limit the two delta functions ${\delta}(\tilde{A}_a)$ and ${\delta}(\hat{A}_a)$ and as a consequence the determinant reduces to

\begin{eqnarray}
\frac{1}{2}Tr_8TR\log{\Omega}^1&=&\frac{1}{2}Tr_8TR\log P^T({\Omega}+ M^2m_T^2)P^T~,~M{\longrightarrow}\infty.
\end{eqnarray}
In above it is consistent to neglect the mass term $M^2m_T^2P^T$ since in the large mass limit $M{\longrightarrow}\infty$ this term  is subleading
as we have discussed.
The eigenvalues of the operator $Adt_c{\cal L}_c$ were computed in the appendix. We found that the second term in $\Omega$ ( which is proportional
 to ${\delta}_{ab}$ ) and the third term ( which is proportional to $(Adt_c{\cal L}_c)_{ab}$ ) can be neglected in the large $n$ limit compared to ${\cal L}_c^2{\delta}_{ab}$. For example the eigenvalues of ${\cal L}_c^2$ are given by $\frac{p^2+2p}{3}$ with $p=0,...,n$ whereas the eigenvalues of $Adt_c{\cal L}_c$ are found to be at most linear in $p$ and hence in the large $n$ limit ( where large values of $p$ which are of the odrer of $n$ are expected to contribute the most ) we can make the approximation
\begin{eqnarray}
{\Omega}_{ab}&{\simeq}&{\phi}^2\bigg({\cal L}_c^2{\delta}_{ab}-\frac{2g^2M^2}{3n^2}\bigg({\cal L}_c^2+\frac{1}{16}-(Adt_c{\cal L}_c-\frac{1}{4})^2\bigg)_{ba}\bigg)+....
\end{eqnarray}
Thus the quantum effective potential is
\begin{eqnarray}
V_{M\longrightarrow \infty}\equiv
\frac{{\Gamma}_1[D]}{6N^2}&=&\frac{S_1[D]}{6N^2}+\frac{1}{6N^2}\bigg(\frac{1}{2}Tr_8TR\log({\phi}^2P^T)-TR\log({\phi}^2)\bigg)+....
\end{eqnarray}
The last term comes from the ghost contribution. The final result is
\begin{eqnarray}
{V}_{M\longrightarrow \infty}
&=&\frac{2}{3n^2g^2}\bigg[\frac{1}{4}{\phi}^4-\frac{1}{2}{\phi}^3+\frac{1}{4}{\phi}^2+\frac{g^2\beta}{9}({\phi}^2-1)^2+\frac{g^2M^2}{27}({\phi}^2-\phi)^2\bigg]+\frac{1}{3}\log\phi.
\end{eqnarray}
The calculation of the  critical values in terms of the mass parameters $\hat{M}^2=g^2M^2$\footnote{This combination is the correct definition of the mass parameter in these models which should be used from the start. } and $\gamma$ is done in the same
way as before  and it yields the following equations. The critical radius occurs at the solutions of the equation
\begin{eqnarray}
\big[1+\frac{4\hat{M}^2}{9}(\gamma +\frac{1}{3})\big]{\phi}_*^2-\frac{9}{8}\big[1+\frac{4\hat{M}^2}{27}\big]{\phi}_*+\frac{1}{4}-\frac{\hat{M}^2}{9}(2\gamma -\frac{1}{3})=0.\label{lkj}
\end{eqnarray}
In the limit $M{\longrightarrow}\infty$ we get the solution
\begin{eqnarray}
{\phi}_{*}{\longrightarrow}\frac{9{+}\sqrt{81+64(1+3\gamma)(1-6\gamma)}}{16(1+3\gamma)}~,~M{\longrightarrow}\infty.\label{phi3}
\end{eqnarray}
The choice of the plus sign instead of the minus sign is so that when $\gamma$ goes to zero ( in other words $\beta {\longrightarrow}0$)
 this solution will reduce to the first equation of (\ref{crit1}). This agreement is due to the fact that the limit $\gamma{\longrightarrow}0$
  is formally equivalent to the limit $M{\longrightarrow}0$ ( since $\gamma=\beta M^2$ ). Indeed for very small values of $M$ we get the potential
\begin{eqnarray}
{V}_{M\longrightarrow 0}
&=&\frac{2}{3n^2g^2}\bigg[\frac{1}{4}{\phi}^4-\frac{1}{2}{\phi}^3+\frac{1}{4}{\phi}^2+\frac{g^2\beta}{9}({\phi}^2-1)^2+\frac{g^2M^2}{27}({\phi}^2-\phi)^2\bigg]+\log\phi.
\end{eqnarray}
This will also lead to the equation (\ref{lkj}) which for $M{\longrightarrow}0$ admits the solution given by the first equation of (\ref{crit1}).

The critical value of the coupling constant $g_*$ ( or equivalently $\bar{\alpha}_*$ ) is given on the other hand by the equation
\begin{eqnarray}
\frac{n^2g_*^2}{2}=\frac{1}{\bar{\alpha}_*^4}=\frac{1}{2}{\phi}_*^2\bigg[\frac{3}{4}(1+\frac{4\hat{M}^2}{27}){\phi}_*-\frac{1}{2}+\frac{2\hat{M}^2}{9}(2\gamma -\frac{1}{3})\bigg].
\end{eqnarray}
Hence in the limit $M{\longrightarrow}\infty$ we get the behavior
\begin{eqnarray}
\bar{\alpha}_*^4=\frac{18}{\hat{M}^2{\phi}_*^2({\phi}_*+4\gamma-\frac{2}{3})+\frac{9}{4}{\phi}_*^2(3{\phi}_*-2)}{\longrightarrow}\frac{18}{\hat{M}^2{\phi}_*^2({\phi}_*+4\gamma-\frac{2}{3})}.\label{pre15}
\end{eqnarray}
The equation of motion $\frac{\partial V_{M{\longrightarrow}\infty}}{\partial{\phi}}=0$ could admit in general four real
solutions where the one with the least energy can be identified with the radius of
fuzzy ${\bf CP}^2_n$.
This  solution is found to be very close to $1$.
However this is only true up to an upper value of the gauge coupling constant $g$ ( or equivalently a lower bound of $\bar{\alpha}$ )
 for every fixed value of $M$  beyond which the equation of motion ceases to have any real solutions. At this value the fuzzy ${\bf CP}^2_n$
  collapses under the effect of quantum fluctuations and we cross   to a pure matrix phase. As the mass $M$ is sent to infinity it is more difficult
  to reach the matrix phase and hence the presence of the mass makes the fuzzy ${\bf CP}^2_n$ solution $D_a=\phi T_a$ more stable. In
fact when $M^2{\longrightarrow}{\infty}$ the critical value $\bar{\alpha}_{*}$
approaches zero.


\paragraph{The transition ${\bf CP}^2{\longrightarrow}{\bf S}^2$}

We   repeat the large mass analysis for the model $S_0$. In other words we add the potential $V_0$ to the action $S_0$ and study the effective
potential when $M,\beta{\longrightarrow}\infty$. The interest in this action lies in the fact that it admits ( at least for $M=\beta =0$ ) a
 fuzzy sphere solution and hence we can contemplate a transition ( at the level of the phase diagram ) between fuzzy ${\bf CP}^2_n$ and fuzzy ${\bf S}^2_N$
  when we take the limit $M,\beta{\longrightarrow}0$. As before we consider fuzzy ${\bf CP}^2_n$ configurations $D_a=\phi T_a$, $a=1,...,8$.
  For $S_0+V_0$ ( in other words non-zero values of $M$ and $\beta$ ) these configurations are in fact  the true vacuum as we have discussed previously.
   When $V_0=0$ the fuzzy ${\bf S}^2_N$ configurations become the true minimum. The calculation of the quantum corrections with non-zero $V_0$ is exactly
    identical to what we have done in the previous paragraphs and thus we end up with the effective potential

\begin{eqnarray}
{V}_{M\longrightarrow \infty}
&=&\frac{2}{3n^2g^2}\bigg[\frac{1}{4}{\phi}^4-\frac{1}{3}{\phi}^3+\frac{g^2\beta}{9}({\phi}^2-1)^2+\frac{g^2M^2}{27}({\phi}^2-\phi)^2\bigg]+\frac{1}{3}\log\phi.
\end{eqnarray}
In the large $M$ limit we get the same critical value
(\ref{phi3}). The critical value of $g$ ( or equivalently
$\bar{\alpha}$ ) is found on the other hand to be given by
\begin{eqnarray}
\bar{\alpha}_*^4=\frac{2}{n^2g_*^2}=\frac{18}{\hat{M}^2{\phi}_*^2({\phi}_*+4\gamma-\frac{2}{3})+\frac{9}{2}{\phi}_*^3}.\label{pre16}
\end{eqnarray}
So again in the large mass limit the fuzzy ${\bf CP}^2_n$ phase is stable even for the model $S_0$.

However we know from our previous discussion that in the limit $M{\longrightarrow}0$ the minimum of the model $S_0$ should tend to the fuzzy sphere
 solutions. Thus it is important to   consider also the fuzzy sphere configurations $
D_i=\phi T_i~,~i=1,2,3~,D_{\alpha}=0~,~\alpha =4,5,6,7,8$. The classical potential in these configurations becomes
\begin{eqnarray}
\frac{S_0[D]}{6N^2}
&=&\frac{1}{12g^2}\bigg[\frac{1}{4}{\phi}^4-\frac{1}{3}{\phi}^3+\frac{\hat{M}^2\gamma c_2}{2n^2}({\phi}^2-\frac{|n|^2}{c_2})^2+\frac{\hat{M}^2}{2}\bigg(\bigg(\frac{2n+3}{6n^2}\bigg)^2{\phi}^2+\frac{c_2}{3n^4}{\phi}^4\bigg)\bigg].
\end{eqnarray}
The quantum corrections in the limit $M{\longrightarrow}0$ should be given by (\ref{enarray1}) and (\ref{enarray2}). We get then the effective potential
\begin{eqnarray}
V_{M\longrightarrow 0}=\frac{{\Gamma}_0[D]}{6N^2}&=&\frac{1}{12g^2}\bigg[\frac{1}{4}{\phi}^4-\frac{1}{3}{\phi}^3+\frac{\hat{M}^2\gamma c_2}{2n^2}({\phi}^2-\frac{|n|^2}{c_2})^2+\frac{\hat{M}^2}{2}\bigg(\bigg(\frac{2n+3}{6n^2}\bigg)^2{\phi}^2+\frac{c_2}{3n^4}{\phi}^4\bigg)\bigg]\nonumber\\
&+&
\log\phi.
\end{eqnarray}
The critical values are
\begin{eqnarray}
{\phi}_*=\frac{3}{4}-\frac{4\hat{M}^2}{3}\bigg[\frac{9\gamma
c_2}{8n^2}-\frac{\gamma
}{3}(1+\frac{3}{n})+\frac{3c_2}{8n^4}+\frac{1}{2}\bigg(\frac{2n+3}{6n^2}\bigg)^2\bigg]+O(\hat{M}^4).
\end{eqnarray}
\begin{eqnarray}
\bar{\alpha}_*^4=\frac{2}{n^2g_*^2}=\frac{96}{n^2{\phi}_*^3}\frac{1}{1+\frac{4\hat{M^2}}{3{\phi}_*}\big[\gamma
(1+\frac{3}{n})-\frac{3}{2}\big(\frac{2n+3}{6n^2}\big)^2\big]}{\longrightarrow}\frac{256\gamma}{3}\hat{M}^2+O(\hat{M}^4).\label{smallmass}
\end{eqnarray}
So when ${M}{\longrightarrow}0$ this $\bar{\alpha}_*$ goes to zero which is consistent with the result (\ref{763}). 
This
 equation tell us how we actually approach this critical value $\bar{\alpha}_*=0$. 

The intersection of this equation with (\ref{pre16}) gives a
  one-loop estimation  of the value $\hat{M}_T$ at which  the vacuum of the model $S_0$ goes from a fuzzy sphere ${\bf S}^2_N$ to a fuzzy ${\bf CP}^2_n$ as
   we increase the mass parameter $\hat{M}$. Equivalently the intersection point occurs at  the value $\hat{M}_T$ at which  the vacuum of the model $S_0$ goes
   from a fuzzy ${\bf CP}^2_n$ to a fuzzy sphere ${\bf S}^2_N$ as we decrease $\hat{M}$.
\section{Conclusion}

In this article we have studied the one-loop effective potential for two models of $U(1)$ gauge theory on fuzzy ${\bf CP}^2_n$. The first model
is given by the action $S_0+V_0$ and the second model is given by the action $S_1$. Each model is characterized by $3$ parameters. $i)$ the gauge
coupling constant $g^2$ or equivalently $\bar{\alpha}^4=\frac{2}{g^2n^2}$, $ii)$ the mass $M$ of the normal components of the $8-$dimensional gauge
field and $iii)$ the parameter $\beta=M^2\gamma$ which gives an extra mass for the normal component in the direction $Adt_8$. The term in the action
proportional to $\beta$ is not needed in the classical theory while in the quantum theory the parameter $\beta$ must be  in the range (\ref{range0}).
 The order parameter ( the variable ) of the effective potential is the radius of  the fuzzy ${\bf CP}^2_n$ and thus by studying the stability of this
 potential we can test the stability of the space as a whole against the effect of quantum fluctuations of the gauge field theory.
 The second term in the effective potential ( the log term )  is not convex which
implies that there is a competition between the classical potential
and the logarithmic term which  depends on the values of $M$ and
$g$. The  parameter $\beta$ plays no further role at this stage. We found that there exists values
of the gauge coupling constant $g$ and the mass $M$ for which the
 fuzzy ${\bf CP}^2_n$ solutions are
not stable. This instability is believed to be related to ( or is a reflection of )  the perturbative UV-IR mixing phenomena of the quantum gauge field
theory. The connection between the two effects was established explicitly for the case of the lower dimensional coadjoint orbit $SU(2)/U(1)$ which is the
 case of the fuzzy sphere \cite{ref}. See also \cite{NCFQED}.

The phase structure of the $U(1)$ models on fuzzy ${\bf CP}^2_n$ which are studied in this article reads as follows.

\paragraph{The model $S_1$}. This is the correct model which describes  $U(1)$
gauge theory in the continuum limit at least classically. The
minimum of the model ( for non-zero potential ) can only be  fuzzy
${\bf CP}^2_n$. There are two phases. In the fuzzy ${\bf CP}^2_n$
phase  we have a $U(1)$ gauge theory on fuzzy ${\bf CP}^2_n$
whereas in the matrix phase the fuzzy ${\bf CP}^2_n$
configurations $D_a=\phi T_a$ collapse and we end up with a $U(N)$
gauge theory on a single point. We have described in this article
the qualitative behavior of a first order phase transition which
occurs between these two regions of the phase space. However it is
obvious from the critical line (\ref{pre16}) that when the mass
$M$ of the four normal scalar
 components of the $8-$dimensional gauge field on  fuzzy ${\bf CP}^{2}_{n}$ goes to infinity it
is more difficult to reach the transition line. In this limit the fuzzy ${\bf CP}^2_n$ phase dominates while the matrix phase shrinks to zero.
Therefore we can say
that we have a nonperturbative  regularization of
$U(1)$ gauge theory on fuzzy ${\bf CP}^{2}_n$.

\paragraph{The model $S_0+V_0$}. This is a string-theory-inspired gauge model which does not go in the continuum limit to the usual $U(1)$ gauge
 theory on ${\bf CP}^2$ even classically. Indeed it can be shown that it contains in the continuum limit ( in addition to the usual Yang-Mills term ) a Chern-Simons-like term  ( see the appendix). However this model has a more interesting  phase structure since it allows for the ( quantum ) transitions between
  fuzzy ${\bf S}^2_N$  and fuzzy ${\bf CP}^2_n$ . The main reason behind this remarkable feature lies in the fact that when $M{\longrightarrow}\infty$ the
   absolute minimum of the model is  the fuzzy ${\bf CP}^2_n$ configurations $D_a=\phi T_a$ whereas in the limit $M{\longrightarrow}0$ the absolute minimum
   of the model is the fuzzy sphere configurations $D_i=\phi T_i,i=1,2,3$ and $D_{\alpha}=0,\alpha =4,5,6,7,8$. The phase diagram of this model with the
    particular value $\gamma=\frac{1}{12}$ is plotted in figure $1$ for illustration . The phase diagram consists of $3$ phases.

\underline{$1)$ The fuzzy ${\bf CP}^2_n$ phase}: This is the region with  $\hat{M}{\geq}\hat{M}_T$ and above the line (\ref{pre16}) where the absolute minimum of the model is  the fuzzy ${\bf CP}^2_n$
 configurations $D_a=\phi T_a$ and where the field theory is some  $U(1)$ gauge theory on fuzzy ${\bf CP}^2_n$. Recall that $\hat{M}_T$ is the
 value of the mass parameter $\hat{M}$ at which the two curves (\ref{pre16}) and (\ref{smallmass}) intersect. The fuzzy ${\bf CP}^2_n$  phase dominates
 the phase diagram when $\hat{M}{\longrightarrow}\infty$. 

\underline{$2)$ The matrix phase}: This  phase  shrinks to zero when $\hat{M}{\longrightarrow}\infty$.
  This occurs at the points of the phase diagram which are below the two lines (\ref{pre16}) and (\ref{smallmass}). 

\underline{$3)$ The fuzzy ${\bf S}^2_N$ phase}: These  are the points which have $\hat{M}{\leq}\hat{M}_T$ and which are   above the line (\ref{smallmass})
    where the absolute minimum of the model is  the fuzzy ${\bf S}^2_N$ configurations $D_i=\phi T_i,i=1,2,3$, $D_{\alpha}=0,\alpha =4,5,6,7,8$
     and where the field theoy is  a $U(1)$ gauge theory on fuzzy ${\bf S}^2_N$ with complicated coupling to $6$ adjoint scalars. 


Generalization to higher gauge groups $U(k)$ with and without fermions should be
straightforward if we are only interested in the effective potential and topology change. 
Similarly generalization to higher coadjoint orbits ${\bf CP}^d=SU(N)/U(N-1)$
with $d=N-1$ should also be straightforward since  the corresponding actions will  be exactly of the same form as $S_0$ and  $S_1$   and  only we need to work with the group theory of $SU(N)$'s instead of $SU(3)$.
In particular we expect that there will more possibilities for topology change in higher coadjoint orbits which relate to the fact that the group $SU(N)$ contains
besides $SU(2)$ the groups $SU(3)$, $SU(4)$ and many others ( for high enough $N$ ) as subgroups. Thus we may see transitions
like ${\bf CP}^d{\longrightarrow}{\bf S}^2$, ${\bf CP}^d{\longrightarrow}{\bf CP}^2$, ${\bf CP}^d{\longrightarrow}{\bf CP}^3$,... as well as
 transitions between  the subspaces ${\bf S}^2$, ${\bf CP}^2$, ${\bf CP}^3$,.. and transitions from and to the  matrix ( single point ) phase.

\begin{figure}
\begin{center}
\includegraphics[width=7cm,angle=-90]{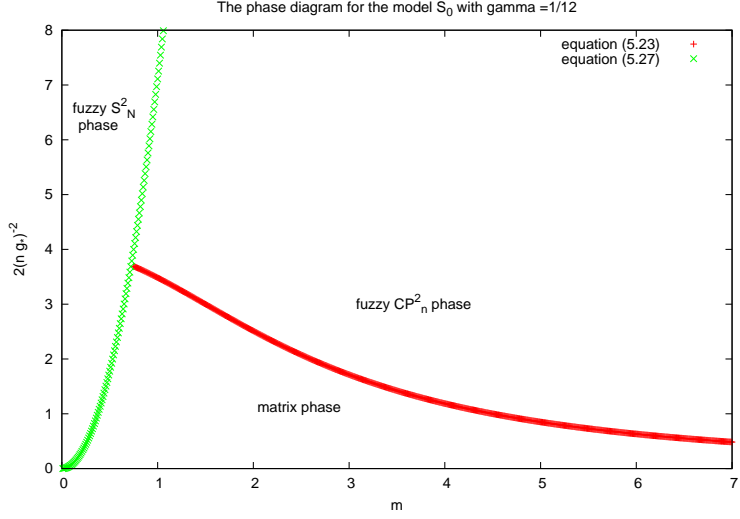}
\caption{The phase diagram for the model $S_0$ with $\gamma=\frac{1}{12}$. In this case ${\phi}_*=1$ and hence the large mass expansion (\ref{pre16})
 becomes $\bar{\alpha}_*^4=\frac{108}{4m^2+27}$ while the small mass expansion (\ref{smallmass}) is $\bar{\alpha}_*^4=\frac{64}{9}m^2$ with $m=\hat{M}$. The intersection point occurs at $\hat{M}_T$. This looks like a triple point. }
\end{center}\label{phase}
\end{figure}
\paragraph{Acknowledgements}
The work of  D.D is
 supported by the College of Science-Research Center Project No:
 phys/2006/04.The work of D.Dou is also 
 supported in part by the associate scheme of Abdus Salam ICTP.
The work of Badis Ydri is supported by a Marie Curie Fellowship
from The Commission of the European Communities ( The Research
Directorate-General ) under contract number MIF1-CT-2006-021797.
B.Y would like also to thank the staff at Humboldt-Universitat zu
Berlin for their help and support. In particular he would like to
thank Michael Muller-Preussker and Wolfgang Bietenholz.
\appendix
\section{The one-loop effective action and effective potential for zero mass}
First we can compute explicitly the following classical actions
\begin{eqnarray}
Tr_NF_{ab}^2=Tr_N\bigg(-[D_a,D_b]^2+4if_{abc}D_aD_bD_c+3D_a^2\bigg).
\end{eqnarray}
\begin{eqnarray}
Tr_N{\Phi}^2=\frac{1}{n^2}Tr_N\bigg[(D_a^2)^2-\frac{2}{3}n(n+3)D_a^2+\frac{1}{9}n^2(n+3)^2\bigg].
\end{eqnarray}
\begin{eqnarray}
Tr_N{\Phi}_a^2&=&\frac{1}{n^2}Tr_N\bigg[\frac{1}{3}D_aD_bD_aD_b+\frac{1}{6}f_{aa^{'}c}f_{bb^{'}c}\{D_a,D_b\}\{D_{a^{'}},D_{b^{'}}\}
+\bigg(\frac{2n+3}{6}\bigg)^2D_a^2\nonumber\\
&-&\frac{2n+3}{3}d_{abc}D_aD_bD_c\bigg].
\end{eqnarray}
In above we have used the identity (\ref{24}) and the identity
\begin{eqnarray}
f_{abc}f_{abd}=3{\delta}_{cd}.
\end{eqnarray}
We can compute ( with $F_{ab}^{(0)}=F_{ab}-i[A_a,A_b]$ )
\begin{eqnarray}
S_0=\frac{1}{2g^2}Tr_NF_{ab}^2-\frac{1}{6g^2}f_{abc}Tr_NA_cF_{ab}-\frac{1}{12g^2}f_{abc}Tr_NA_cF_{ab}^{(0)}.
\end{eqnarray}
Next we  will study the quantization of the action
\begin{eqnarray}
{S}_i[D,J]&=&S_i[D]+Tr_NJ_aD_a~,~S_i[D]=\bar{S}_i[D]+\bar{V}_i[D]
\end{eqnarray}
where
\begin{eqnarray}
\bar{S}_i[D]&=&\frac{1}{g^2}Tr_N\bigg(-\frac{1}{4}[D_a,D_b]^2+i{\alpha}_i f_{abc}D_aD_bD_c\bigg)\nonumber\\
\bar{V}_i[D]&=&{\rho}_i Tr_N\Phi +V_i={\rho}_i Tr_N\Phi +{\beta}_i Tr_N{\Phi}^2 +M_i^2 Tr_N{\Phi}_c^2.\label{okok}
\end{eqnarray}
For $S_0=\bar{S}_0+\bar{V}_0$ we have ${\alpha}_0=\frac{2}{3}$, ${\rho}_0={\beta}_0=M_0=0$ while for $S_1=\bar{S}_1+\bar{V}_1$ we have ${\alpha}_1=1$, ${\rho}_1 =\frac{3n}{4g^2}$, ${\beta}_1=\beta$ and $M_1=M$. $J_a$ is a source.

We adopt the background field method to the problem of quantization of this model. We write $D_a=B_a+Q_a$ where $B_a$ is the background field and $Q_a$ is the fluctuation field. We will fix the gauge by adding to the action the gauge-fixing and Fadeev-Popov terms, viz

\begin{eqnarray}
S_{\rm g.fixing}+S_{\rm gh}&=&-\frac{1}{2g^2}\frac{1}{\xi}Tr_N[B_a,Q_a]^2-Tr_N[B_a,b^+][B_a,b].
\end{eqnarray}
We compute
\begin{eqnarray}
Tr_N[D_a,D_b]^2&=&Tr_N\bigg([B_a,B_b]^2+2[B_a,B_b][Q_a,Q_b]+2[B_a,Q_b]^2+2[B_a,Q_b][Q_a,B_b]\nonumber\\
&+&4[B_a,B_b][B_a,Q_b]+O(Q^3)\bigg).
\end{eqnarray}
And
\begin{eqnarray}
if_{abc}Tr_ND_aD_bD_c&=&if_{abc}Tr_NB_aB_bB_c+3if_{abc}Tr_NB_aB_bQ_c+3if_{abc}Tr_NQ_aQ_bB_c.
\end{eqnarray}
We find ( by using the identity $Tr_N[B_a,Q_b][B_b,Q_a]=Tr_N[B_a,Q_a]^2-Tr_N[B_a,B_b][Q_a,Q_b]$ ) the following expression
\begin{eqnarray}
\bar{S}_i[D]&=&\bar{S}_i[B]+\frac{i}{g^2}Tr_N\bigg[B_{ab}-(1-\frac{3{\alpha}_i}{2})f_{abc}B_c,B_a\bigg]Q_b-\frac{1}{2g^2}Tr_N\bigg([B_a,Q_b]^2\nonumber\\
&-&[B_a,Q_a]^2+2iQ_a\bigg[B_{ab}-(1-\frac{3{\alpha}_i}{2})f_{abc}B_c,Q_b\bigg]\bigg)+O(Q^3)\nonumber\\
&=&\bar{S}_i[B]+\frac{i}{g^2}Tr_N\bigg[B_{ab}-(1-\frac{3{\alpha}_i}{2})f_{abc}B_c,B_a\bigg]Q_b+\frac{1}{2g^2}Tr_NQ_a\bigg({\cal B}_c^2{\delta}_{ab}\nonumber\\
&-&{\cal B}_a{\cal B}_b-2i{\cal B}_{ab}+2i(1-\frac{3{\alpha}_i}{2})f_{abc}{\cal B}_c\bigg)Q_b+O(Q^3).\label{47}
\end{eqnarray}
$B_{ab}$ is the curvature of the background curvature $B_a$, in
other words $B_{ab}=i[B_a,B_b]+f_{abc}B_c$. Remark also how this
action simplifies for $S_0$, i.e for ${\alpha}_0 =\frac{2}{3}$.
This ``technical'' simplification is a major advantage in
considering $S_0$ instead of $S_1$. Given a matrix  $O$ the
operator ${\cal O}$ is given by ${\cal O}(..)=[O,..]$, for example
${\cal B}_a(Q_a)=[B_a,Q_a]$.

We can also compute
\begin{eqnarray}
{\rho}_i Tr_N\Phi ={\rho}_i Tr_N\Psi +\frac{2{\rho}_i}{n}Tr_NB_bQ_b+\frac{{\rho}_i}{n}Tr_NQ_a^2.
\end{eqnarray}
\begin{eqnarray}
{\beta}_i Tr_N{\Phi}^2&=&{\beta}_i Tr_N{\Psi}^2+\frac{2{\beta}_i}{n}Tr_N\{B_b,{\Psi}\}Q_b+\frac{{\beta}_i}{n^2}Tr_NQ_a\bigg(-{\cal B}_{a}{\cal B}_{b}+4B_{a}B_{b}+2n{\delta}_{ab}\Psi\bigg)Q_b\nonumber\\
&+& O(Q^3).
\end{eqnarray}
\begin{eqnarray}
M_i^2 Tr_N{\Phi}_c^2&=&M_i^2 Tr_N{\Psi}_c^2+\frac{2M_i^2}{n}Tr_N\bigg(d_{abc}\{B_b,{\Psi}_c\}-\frac{2n+3}{6}{\Psi}_b\bigg)Q_b\nonumber\\
&+&\frac{M_i^2}{n^2}Tr_NQ_a\bigg(-d_{aa^{'}c}d_{bb^{'}c}{\cal B}_{a^{'}}{\cal B}_{b^{'}}+4d_{aa^{'}c}d_{bb^{'}c}B_{a^{'}}B_{b^{'}}-2(\frac{2n+3}{3})d_{abc}B_c\nonumber\\
&+&2nd_{abc}{\Psi}_c+(\frac{2n+3}{6})^2{\delta}_{ab}\bigg)Q_b +O(Q^3).
\end{eqnarray}
$\Psi$ and ${\Psi}_c$ are the normal scalar fields corresponding
to the background covariant derivative $B_a$, viz
${\Psi}=\frac{1}{n}(B_a^2-|n|^2)$ and
${\Psi}_c=\frac{1}{n}(d_{abc}B_aB_b-\frac{2n+3}{6}B_c)$ where
$|n|^2=\frac{1}{3}n(n+3)$.

Let us now introduce the actions
${S}_i[D,J]=\bar{S}_i[D]+\bar{V}_i[D]+Tr_NJ_aD_a$ and
${S}_i[B,J]=\bar{S}_i[B]+\bar{V}_i[B]+Tr_NJ_aB_a$. By using the
above ingredients we have immediately the result
\begin{eqnarray}
{S}_i[D,J]+S_{\rm g.fixing}+S_{\rm gh}&=&{S}_i[B,J]+Tr_N{\cal J}_b^iQ_b+\frac{1}{2g^2}Tr_NQ_a{\Omega}_{ab}^iQ_b+O(Q^3)\nonumber\\
&+&Tr_Nb^+{\cal B}^2_ab.
\end{eqnarray}
In above ${\cal J}_a^i$ and ${\Omega}_{ab}^i$ are given respectively by

\begin{eqnarray}
{\cal J}_b^i&=&J_b+\frac{i}{g^2}[B_{ab}-(1-\frac{3{\alpha}_i}{2})f_{abc}B_c,B_a]+\frac{2{\rho}_i}{n}B_b\nonumber\\
&+&\frac{2{\beta}_i}{n}\{B_b,\Psi\}
+\frac{2M_i^2}{n}(d_{abc}\{B_b,{\Psi}_c\}-\frac{2n+3}{6}{\Psi}_b)
\end{eqnarray}
\begin{eqnarray}
{\Omega}_{ab}^i&=&{\cal B}_c^2{\delta}_{ab}+(\frac{1}{\xi}-1){\cal B}_a{\cal B}_b-2i{\cal B}_{ab}+2i(1-\frac{3{\alpha}_i}{2})f_{abc}{\cal B}_c+\frac{2g^2{\rho}_i}{n}{\delta}_{ab}\nonumber\\
&+&\frac{2g^2{\beta}_i}{n^2}\bigg(-{\cal B}_{a}{\cal B}_{b}+4B_{a}B_{b}+2n{\delta}_{ab}\Psi\bigg)+\frac{2g^2M_i^2}{n^2}\bigg(-d_{aa^{'}c}d_{bb^{'}c}{\cal B}_{a^{'}}{\cal B}_{b^{'}}\nonumber\\
&+&4d_{aa^{'}c}d_{bb^{'}c}B_{a^{'}}B_{b^{'}}-2(\frac{2n+3}{3})d_{abc}B_c+2nd_{abc}{\Psi}_c
+(\frac{2n+3}{6})^2{\delta}_{ab}\bigg).
\end{eqnarray}
In the following we will assume that the background fields $B_a$ satisfy the equations of motion ${\cal J}_b^i=0$ and we will choose the gauge ${\xi}^{-1}=1+\frac{2g^2{\beta}_i}{n^2}$. We then obtain
\begin{eqnarray}
{S}_i[D,J]+S_{\rm g.fixing}+S_{\rm gh}&=&{S}_i[B,J]+\frac{1}{2g^2}Tr_NQ_a{\Omega}_{ab}^iQ_b+O(Q^3)+Tr_Nb^+{\cal B}^2_ab.
\end{eqnarray}
The fluctuation fields $Q_a$ and the ghosts can be integrated out
since they are Gaussian and one obtains therefore the effective
action
\begin{eqnarray}
{\Gamma}_i[B,J]&=&{S}_i[B,J]+\frac{1}{2}Tr_8TR\log{\Omega}^i_{ab}-TR\log{\cal B}_a^2.
\end{eqnarray}
where
\begin{eqnarray}
{\Omega}_{ab}^i&=&{\cal B}_c^2{\delta}_{ab}-2i{\cal B}_{ab}+2i(1-\frac{3{\alpha}_i}{2})f_{abc}{\cal B}_c+\frac{2g^2{\rho}_i}{n}{\delta}_{ab}+\frac{2g^2{\beta}_i}{n^2}\bigg(4B_{a}B_{b}+2n{\delta}_{ab}\Psi\bigg)\nonumber\\
&+&\frac{2g^2M_i^2}{n^2}\bigg(-d_{aa^{'}c}d_{bb^{'}c}{\cal B}_{a^{'}}{\cal B}_{b^{'}}+4d_{aa^{'}c}d_{bb^{'}c}B_{a^{'}}B_{b^{'}}-2(\frac{2n+3}{3})d_{abc}B_c+2nd_{abc}{\Psi}_c\nonumber\\
&+&(\frac{2n+3}{6})^2{\delta}_{ab}\bigg).
\end{eqnarray}

Now we compute the effective potential on fuzzy ${\bf CP}^2$ for $V_0=0$.
For the configuration (\ref{cp2-conf}) we consider the following two cases

\paragraph{Case $1$} For $S_0$ we have ${\alpha}_0 =\frac{2}{3}$ and ${\rho}_0 =0$ and hence
\begin{eqnarray}
&&{\Gamma}_0[D]=\frac{3|n|^2}{g^2}\bigg[\frac{1}{4}{\phi}^4-\frac{1}{3}{\phi}^3\bigg]+\frac{1}{2}Tr_8TR\log{\Omega}^0_{ab}-TR\log{\phi}^2{\cal L}_a^2\nonumber\\
&&{\Omega}^0_{ab}={\phi}^2{\cal L}_c^2{\delta}_{ab}-{\phi}(\phi -1)\bigg[{\cal J}_c^2-{\cal L}_c^2-(Adt_c)^2\bigg]_{ab}~,~\vec{\cal J}=\vec{\cal L}+{Ad}\vec{t}.
\end{eqnarray}
The total $SU(3)$ angular momentum $\vec{\cal J}$ corresponds to the tensor product of the irreducible representations $(p,p)$ where $p=0,...,n$ ( corresponding to $\vec{\cal L}$ ) with the adjoint representation $(1,1)$ ( corresponding to $Ad\vec{t}$ ). By using Young tableaux we obtain the decomposition
\begin{eqnarray}
(p,p){\otimes}(1,1)&=&(p+1,p+1)\oplus( p-1,p-1)\oplus (p,p)\oplus(p,p)\oplus (p-1,p+2) \oplus (p+2,p-1)\nonumber\\
& \oplus & (p-2,p+1)\oplus (p+1,p-2).
\end{eqnarray}
The dimension of an irreducible representation $(n_1,n_2)$ of
$SU(3)$ is $d(n_1,n_2)=\frac{1}{2}(n_1+1)(n_2+1)(n_1+n_2+2)$ and
the quadratic Casimir is
$c_2(n_1,n_2)=\frac{1}{3}(n_1^2+n_1n_2+3n_1+3n_2+n_2^2)$. Thus we
can immediately compute
\begin{eqnarray}
{\cal J}_c^2-{\cal L}_c^2-(Adt_c)^2={\cal J}_c^2-(p^2+2p)-3={\rm diag}({2p},
-(2p+4),-3,0,-p-3). \label{c2p}
\end{eqnarray}
In this diagonal matrix the dimensions of the first block is $d(p+1,p+1)=(p+2)^3$, the second block is $d(p-1,p-1)=p^3$, the third block is $2d(p,p)=2(p+1)^3$, the $4$th block is $d(p-1,p+2)+d(p+2,p-1)=2d(p-1,p+2)=p(p+3)(2p+3)$ and the $5$th block is $d(p-2,p+1)+d(p+1,p-2)=2d(p-2,p+1)=(p-1)(p+2)(2p+1)$.

It is  obvious from the above equation (\ref{c2p}) that the second term in ${\Omega}$ is at most linear in $p$ while the first term is quadratic and hence in the large $n$ limit ( where large values of $p$ which are of the odrer of $n$ are expected to contribute the most ) we can make the approximation ${\Omega}_{ab}={\phi}^2{\cal L}_c^2{\delta}_{ab}$

Thus the quantum effective potential is
\begin{eqnarray}
{\Gamma}_0[D]&=&S_0[D]+\frac{1}{2}Tr_8TR\log({\phi}^2{\cal L}_a^2{\bf 1}_8)-TR\log({\phi}^2{\cal L}_a^2)\nonumber\\
&=&\frac{3|n|^2}{g^2}\bigg[\frac{1}{4}{\phi}^4-\frac{1}{3}{\phi}^3\bigg]+\frac{1}{2}(8)(N^2)\log{\phi}^2-(N^2)\log{\phi}^2+{\rm constant}\nonumber\\
&=&\frac{3|n|^2}{g^2}\bigg[\frac{1}{4}{\phi}^4-\frac{1}{3}{\phi}^3\bigg]+6N^2\log{\phi}+{\rm constant}.
\end{eqnarray}
Recall that $N=\frac{1}{2}(n+1)(n+2)$ and
$|n|^2=\frac{1}{3}n(n+3)$ and hence
\begin{eqnarray}
V_{\rm eff}=\frac{\Gamma[B]}{6N^2}
&=&\frac{2}{3n^2g^2}\bigg[\frac{1}{4}{\phi}^4-\frac{1}{3}{\phi}^3\bigg]+\log{\phi}+{\rm constant}.\label{rmconstant}
\end{eqnarray}

\paragraph{Case $2$} For $S_1$ we have ${\alpha}_1 =1$ and ${\rho}_1 =\frac{3n}{4g^2}$ and hence

\begin{eqnarray}
{\Omega}^1_{ab}&=&\bigg({\phi}^2{\cal L}_c^2+\frac{3}{2}\bigg){\delta}_{ab}+2i\phi(\phi -\frac{3}{2})f_{abc}{\cal L}_c.
\end{eqnarray}
The only  difference ( as far as this quantum correction is concerned ) with case  $1$  is that we have now a shifted Laplacian ${\phi}^2{\cal L}_c^2+\frac{3}{2}$ so the result for the determinant already obtained will not be altered.
We end up thus with the effective potential
\begin{eqnarray}
V_{\rm eff}=\frac{\Gamma[B]}{6N^2}
&=&\frac{2}{3n^2g^2}\bigg[\frac{1}{4}{\phi}^4-\frac{1}{2}{\phi}^3+\frac{1}{4}{\phi}^2\bigg]+\log{\phi}+{\rm constant}.\label{rmconstant1}
\end{eqnarray}

\bibliographystyle{unsrt}

\begin{thebibliography}{99}


\bibitem{thesis}
Badis Ydri, {\it Fuzzy Physics}, PhD thesis (2001), {\it
hep-th/0110006}.
\bibitem{thesis1}
H.Grosse,C.Klim\v cik,P.Pre \v snajder,{\it Commun.Math.Phys.} 180
(1996) 429, {\it Int.J.Theor.Phys.} 35 (1996) 231. D.O'Connor,
{\it Mod.Phys.Lett.} A18 (2003) 2423. C.Klim\v cik, {\it
Commun.Math.Phys.} 199 (1998) 257.
  A.~P.~Balachandran, S.~Kurkcuoglu and S.~Vaidya,''Lectures on fuzzy and fuzzy SUSY physics'', {\it hep-th/0511114} .


\bibitem{CONNES}
A. Connes, {\it Noncommutative Geometry}, Academic Press, London ,
1994 . G. Landi, {\it An introduction to noncommutative spaces and
their geometry}, springer (1997). J. M. Gracia-Bondia, J. C.
Varilly, H. Figueroa, {\sl Elements of Noncommutative Geometry},
Birkhauser (2000).


\bibitem{DENJ2006}
Brian P. Dolan, Idrish Huet, Sean Murray, Denjoe O'Connor, {\it
hep-th/0611209}.

\bibitem{IKKT}
N. Ishibashi, H. Kawai, Y. Kitazawa, A. Tsuchiya, {\it Nucl.Phys.
}B498 (1997) 467-491.

\bibitem{IKKT1}
H. Aoki, S. Iso, H. Kawai, Y. Kitazawa, T. Tada, {\it
Prog.Theor.Phys. }99 (1998) 713-746. H. Aoki, S. Iso, H. Kawai, Y.
Kitazawa, T. Tada, A. Tsuchiya,  {\it Prog.Theor.Phys.Suppl. }134
(1999) 47-83. H. Aoki, N. Ishibashi, S. Iso, H. Kawai, Y.
Kitazawa, T. Tada, {\it Nucl.Phys.} B565 (2000) 176-192.










\bibitem{madore} J.Madore,{\it Class.Quant.Grav.} 9:69-88,1992.
J.Hoppe, MIT PhD thesis (1982). J.Hoppe, S.T.Yau, {\it
Commun.Math.Phys.}195(1998)67-77.

\bibitem{iso}
S.Iso, Y.Kimura, K.Tanaka, K.Wakatsuki, {\it Nucl.Phys.} B604
(2001) 121.






\bibitem{ref} P.Castro-Villarreal , R.Delgadillo-Blando , { Badis Ydri} , {\it Nucl.Phys.B} {\bf 704} (2004) 111-153.

\bibitem{subrata}
T.Azuma , S.Bal, K.Nagao , J.Nishimura,  {\it hep-th/0405277}.




\bibitem{ref11} D.O'Connor, { Badis Ydri},  {\it hep-lat/0606013}, {\it JHEP}11(12006)016.

\bibitem{ref10}
Badis Ydri, {\sl Quantum equivalence of NC and YM gauge theories
in 2D and matrix theory}, {\it hep-th/0701057}.
 { Badis Ydri} , {\sl The one-plaquette model limit of NC gauge theory in 2D}, {\it hep-th/0606206}, to be published in NPB.




\bibitem{ars}
A.Y.Alekseev , A.Recknagel, V.Schomerus, {\it hep-th/0003187} and
{\it hep-th/9812193} .

\bibitem{NCFQED}
 P.Castro-Villarreal , R.Delgadillo-Blando , { Badis Ydri}, {\it JHEP}09 (2005)066.
 R.Delgadillo-Blando , { Badis Ydri}, {\sl Towards Noncommutative Fuzzy QED}, {\it hep-th/0611177}.

\bibitem{paulo}
A.P. Balachandran, S. Kurkcuoglu, {\it
Int.J.Mod.Phys.}A19:3395-3408,2004. Luiz C. de Albuquerque, Paulo
Teotonio-Sobrinho, Sachindeo Vaidya {\it JHEP.} 0410:024,2004.

\bibitem{hoppe}
J.Arnlind, M.Bordemann, L.Hofer, J.Hoppe, H.Shimada, {\it hep-th/0602290}.

\bibitem{CP2}
G.Alexanian, A.P.Balachandran, G.Immirzi, B.Ydri, {\it
J.Geo.Phys.}{42} (2002) 28-53.
\bibitem{cp2}
H.Grosse, A Strohmaier, {\it Lett.Math.Phys.} {48} (1999) 163-179.


\bibitem{brian}
A.P. Balachandran, Brian P. Dolan, J. Lee, X. Martin, Denjoe
O'Connor,~{\it  hep-th/0107099}.



\bibitem{stein}
H.Grosse, H.Steinacker,{\it hep-th/0407089}.



















\end{thebibliography}

\end{document}